\documentclass[rmp,twocolumn,twoside]{revtex4}

\usepackage{dcolumn}
\usepackage{amsmath}
\usepackage{bibentry}
\usepackage{pslatex}

\newcommand{\etal}{{\it{}et~al.}}
\newcommand{\defn}{\textit}

\newcounter{bibnumber}
\newcommand\bibnum[1]{\refstepcounter{bibnumber}%
  \smallbreak\par\noindent\textbf{\arabic{bibnumber}}.\quad\bibentry{#1}.}

\newenvironment{bibblock}{%
  \par\vspace{1.5ex}\small}{\vspace{1.5ex}\par}

\begin{document}

\bibliographystyle{ajprl}
\nobibliography{journals,references}
\newcommand{\enquote}[1]{``#1''}

\title{Complex Systems: A Survey}
\author{M. E. J. Newman}
\affiliation{Department of Physics, University of Michigan, Ann Arbor, MI
48109}
\affiliation{Center for the Study of Complex Systems, University of
Michigan, Ann Arbor, MI 48109}

\begin{abstract}
  A complex system is a system composed of many interacting parts, often
  called agents, which displays collective behavior that does not follow
  trivially from the behaviors of the individual parts.  Examples include
  condensed matter systems, ecosystems, stock markets and economies,
  biological evolution, and indeed the whole of human society.  Substantial
  progress has been made in the quantitative understanding of complex
  systems, particularly since the 1980s, using a combination of basic
  theory, much of it derived from physics, and computer simulation.  The
  subject is a broad one, drawing on techniques and ideas from a wide range
  of areas.  Here I give a survey of the main themes and methods of complex
  systems science and an annotated bibliography of resources, ranging from
  classic papers to recent books and reviews.
\end{abstract}

\maketitle

\section{Introduction}
Complex systems is a relatively new and broadly interdisciplinary field
that deals with systems composed of many interacting units, often called
``agents.''  The foundational elements of the field predate the current
surge of interest in it, which started in the 1980s, but substantial recent
advances in the area, coupled with increasing interest both in academia and
industry, have created new momentum for the study and teaching of the
science of complex systems.

There is no precise technical definition of a ``complex system,'' but most
researchers in the field would probably agree that it is a system composed
of many interacting parts, such that the collective behavior of those parts
together is more than the sum of their individual behaviors.  The
collective behaviors are sometimes also called ``emergent'' behaviors, and
a complex system can thus be said to be a system of interacting parts that
displays emergent behavior.

Classic examples of complex systems include condensed matter systems,
ecosystems, the economy and financial markets, the brain, the immune
system, granular materials, road traffic, insect colonies, flocking or
schooling behavior in birds or fish, the Internet, and even entire human
societies.

Unfortunately, complex systems are, as their name makes clear, complex,
which makes them hard to study and understand.  Experimental observations
are of course possible, though these fall largely within the realm of the
traditional scientific disciplines and are usually not considered a part of
the field of complex systems itself, which is primarily devoted to
theoretical developments.

Complex systems theory is divided between two basic approaches.  The first
involves the creation and study of simplified mathematical models that,
while they may not mimic the behavior of real systems exactly, try to
abstract the most important qualitative elements into a solvable framework
from which we can gain scientific insight.  The tools used in such studies
include dynamical systems theory, game theory, information theory, cellular
automata, networks, computational complexity theory, and numerical methods.
The second approach is to create more comprehensive and realistic models,
usually in the form of computer simulations, which represent the
interacting parts of a complex system, often down to minute details, and
then to watch and measure the emergent behaviors that appear.  The tools of
this approach include techniques such as Monte Carlo simulation and,
particularly, agent-based simulation, around which a community of computer
scientists and software developers has grown up to create software tools
for sophisticated computational research in complex systems.

This review focuses on the methods and theoretical tools of complex
systems, including both the modeling and simulation approaches, although I
also include a short section of references on individual complex systems,
such as economies or ecosystems, which can serve as a concrete foundation
motivating the theoretical studies.

\section{General references}
Complex systems is a relatively young subject area and one that is evolving
rapidly, but there are nonetheless a number of general references,
including books and reviews, that bring together relevant topics in a
useful way.

\subsection{Books}
\label{sec:genbooks}

The first two books listed are elementary and require little mathematics
for their comprehension.  The first, by Mitchell, is recent and aimed at
the popular audience.  The second is older but wider ranging and contains
more technical content.\footnote{Throughout this review each bibliographic
  reference is labeled ``(E)'', ``(I)'', or ``(A),'' to denoted elementary,
  intermediate, or advanced material.  Elementary material requires only a
  modest level of mathematical know\-ledge, intermediate material is
  suitable for readers with a solid grasp of mathematics at the
  undergraduate level, and advanced material is suitable for advanced
  undergraduates or beginning graduate students in mathematics or the
  physical sciences.}
\begin{bibblock}
\bibnum{Mitchell09}\label{bib:mitchell}~(E)
\bibnum{Flake98}~(E)
\end{bibblock}
The following three books are more advanced.  Each covers important topics
in complex systems, though none covers the field comprehensively.  The
authors of the second book are economists and their book has, as a result,
more of a social science flavor.  The book by Mandelbrot is, by now, quite
old, predating ``complex systems'' as a recognized field, but is considered
a classic and very readable, although not all of the ideas it contains have
become accepted.
\begin{bibblock}
\bibnum{Boccara04}~(I)
\bibnum{MP07}\label{bib:MP}~(I)
\bibnum{Mandelbrot83}\label{bib:mandelbrot}~(I)
\end{bibblock}

\section{Examples of complex systems}
\label{sec:individual}
Many individual complex systems are studied intensively within their own
academic fields---ecosystems in ecology, stock markets in finance and
business, and so forth.  It is not the purpose of this paper to review this
subject-specific literature, but this section outlines some of the
literature on specifically complex-systems approaches to individual
systems.

\paragraph*{Physical systems:}
Although they are not always thought of in that way, many physical systems,
and particularly those studied in condensed matter and statistical physics,
are true examples of complex systems.  Physical systems that fall within
the realm of complex systems science include classical condensed matter
systems such as crystals, magnets, glasses, and superconductors;
hydrodynamical systems including classical (Newtonian) fluids, nonlinear
fluids, and granular flows; spatiotemporal pattern formation in systems
like chemical oscillators and excitable media; molecular self-assembly,
including tiling models, biomolecules, and nanotechnological examples;
biophysical problems such as protein folding and the physical properties of
macromolecules; and physical systems that perform computation, including
analog and quantum computers.  It is perhaps in condensed matter physics
that the fundamental insight motivating the study of complex systems was
first clearly articulated, in the classic 1972 article by Anderson:
\begin{bibblock}
  \bibnum{Anderson72}\quad In this paper Anderson points out
  the misconception of basic physical theories, such as quantum mechanics,
  as ``theories of everything.''  Although such theories do, in principle,
  explain the action of the entire universe, the collective behaviors of
  particles or elements in a complex system often obey emergent physical
  laws---like the equation of state of a gas, for instance---that cannot be
  derived easily (or in some cases at all) from the underlying microscopic
  theory.  In other words, there are physical laws at many ``levels'' in
  the phenomenology of the universe, and only one of those levels is
  described by fundamental theories like quantum mechanics.  To understand
  the others, new theories are needed.~(E)
\end{bibblock}
Many of the physicists who have made careers working on complex systems got
their start in condensed matter physics, and an understanding of that field
will certainly help the reader in understanding the ideas and language of
complex systems theory.  Two recent books written by physicists directly
involved in research on complex systems are:
\begin{bibblock}
\bibnum{Sethna06}\quad This book is accompanied by a set of online
  programs and simulations that are useful for explaining and understanding
  some of the concepts.~(A)

\bibnum{Sander09}~(A)
\end{bibblock}
Both are sophisticated treatments, but for the mathematically inclined
reader these books provide a good starting point for understanding physical
theories of complex systems.

\paragraph*{Ecosystems and biological evolution:}
The biosphere, both in its present state and over evolutionary history,
presents an endlessly fascinating picture of a complex system at work.
\begin{bibblock}
\bibnum{SG02}\quad A good introduction, which includes some significant
  mathematical elements, but confines the most challenging of them to
  sidebars.  The authors are a physicist and a biologist, and the
  combination makes for a book that is accessible and relevant to those
  interested in how physics thinking can contribute outside of the
  traditional boundaries of physics.~(I)

\bibnum{Nowak06}\label{bib:nowak}\quad A more technical work that also
  includes an introduction, in the biological arena, to several of the
  areas of complex systems theory discussed later in this paper.~(I)
\end{bibblock}
The following two papers provide useful discussions from the ecology
viewpoint:
\begin{bibblock}
\bibnum{Levin98}~(I)
\bibnum{Holling01}\quad As its title suggests, this article
  provides a comparative review of ecosystems along side economies and
  human societies, from the viewpoint of an ecologist.~(I)
\end{bibblock}
Some classic works in complex systems also fall into the areas of ecology
and evolutionary biology:
\begin{bibblock}
\bibnum{May72}\quad This important early paper applies
  complex systems ideas to the stability of ecosystems, and is a
  significant precursor to more recent work in network theory (see
  Section~\ref{sec:networks}).~(A)
\bibnum{KL87}\quad In this paper, Kauffman and Levin
  described for the first time their NK model, which is now one of the
  standard models of macroevolutionary theory.~(A)
\bibnum{Kauffman95}\label{bib:kauffman}\quad This later book by Kauffman
gives an accessible introduction to the NK model.~(E)
\end{bibblock}

\paragraph*{Human societies:}
Human societies of course have many aspects to them, not all of which are
amenable to study by quantitative methods.  Three aspects of human
societies, however, have proved of particular interest to scientists
working on complex systems: (1)~urban planning and the physical structure
of society, (2)~the social structure of society and social networks, and
(3)~differences between societies as revealed by sociological experiments.
I address the first two of these in this section.  Experimental approaches
are addressed in Section~\ref{sec:gametheory}.

One of the most influential works on urban planning is the 1961 book by
Jacobs which, while predating modern ideas about complex systems, has
nonetheless inspired many of those ideas.  It is still widely read today:
\begin{bibblock}
\bibnum{Jacobs61}~(E)
\end{bibblock}
The following papers provide a sample of recent work on urban societies
viewed as complex systems.  The articles by Bettencourt~\etal, which
address the application of scaling theory to urban environments, have been
particularly influential, although their results are not universally
accepted.  The first is at a relatively high technical level while the
second is a non-technical overview.  I discuss scaling theory in more
detail in Section~\ref{sec:scaling}.
\begin{bibblock}
\bibnum{Batty08}\quad Batty is an architect who has in recent years
  championed the application of complex systems theory in urban planning.
  In this nontechnical article he gives an overview of current ideas,
  drawing on spatial models, scaling, and network theory.~(E)

\bibnum{Batty07}\quad In this book Batty expands widely on
  the topic of his article above.  Although technical, the book is
  approachable and the author makes good use of models and examples to
  support his ideas.~(I)

\bibnum{Bettencourt07}\quad The work of Bettencourt and collaborators on
  the application of scaling theory to the study of urban environments has
  been particularly influential.  They find that a wide variety of
  parameters describing the physical structure of US cities show
  ``power-law'' behavior.  Power laws are discussed further in
  Section~\ref{sec:scaling}.~(A)

\bibnum{BW10}\quad This nontechnical paper discusses the
  motivations and potential rewards of applying complex systems approaches
  to urban planning.~(E)
\end{bibblock}
Turning to social networks, there has been a substantial volume of work on
networks in general by complex systems researchers, which we review in
Section~\ref{sec:networks}, but there is also an extensive literature on
human social networks in sociology, which, while not specifically aimed at
readers in complex systems, nonetheless contains much of interest.  The two
books below are good general references.  The article by Watts provides an
interesting perspective on what complex systems theory has to add to a
field of study that is now almost a hundred years old.
\begin{bibblock}
\bibnum{Scott00}~(I)
\bibnum{WF94}~(A)
\bibnum{Watts04}~(I)
\end{bibblock}

\paragraph*{Economics and markets:}
Markets are classic examples of complex systems, with manufacturers,
traders, and consumers interacting to produce the emergent phenomenon we
call the economy.  Physicists and physics-style approaches have made
substantial contributions to economics and have given rise to the new
subfield of ``econophysics,'' an area of lively current research activity.
\begin{bibblock}
\bibnum{MS99}\quad This book is a standard reference in the area.~(I)
\bibnum{Sornette04}\quad Though it addresses primarily financial markets,
and not economics in general, this highly-regarded book is a good example
of the physics approach to these problems.~(I)
\bibnum{FSS05}\quad An approachable introductory paper that asks what
physics can contribute to our understanding of economic and financial
problems.~(E)
\end{bibblock}
A fundamental debate that has characterized the influence of complex
systems ideas on economics is the debate over the value of the traditional
``equilibrium'' models of mathematical economics, as opposed to newer
approaches based on ideas such as ``bounded rationality'' or on computer
simulation methods.  A balanced overview of the two viewpoints is given by
Farmer and Geanakoplos.
\begin{bibblock}
\bibnum{FG09}~(E)
\end{bibblock}
A number of books have also appeared that make connections between economic
theory and other areas of interest in complex systems.  A good recent
example is the book by Easley and Kleinberg, which draws together ideas
from a range of fields to help illuminate economic behaviors and many other
things in a lucid though quantitative way.
\begin{bibblock}
\bibnum{EK10}\label{bib:kleinberg}~(E)
\end{bibblock}

\paragraph*{Pattern formation and collective motion:}
In two- or three-dimensional space the interactions of agents in a complex
system can produce spatial patterns of many kinds and systems that do this
are seen in many branches of science, including physics
(e.g.,~Rayleigh--B\'enard convection, diffusion limited aggregation),
chemistry (the Belousov--Zhabotinsky reaction), and biology (embryogenesis,
bacterial colonies, flocking and collective motion of animals and humans).
The paper by Turing below is one of the first and best-known efforts to
develop a theory of pattern formation in the context of biological
morphogenesis, and a classic in the complex systems literature.  The book
by Winfree is an unusual and thought-provoking point of entry into the
literature that makes relatively modest mathematical demands of its reader
(and addresses many other topics in addition to pattern formation).
\begin{bibblock}
\bibnum{Turing52}~(A)
\bibnum{Winfree00}~(I)
\end{bibblock}
Collective motions of self-propelled agents, such as road and pedestrian
traffic and animal flocking, have been actively studied using methods from
physics.  Vehicular traffic shows a number of interesting behaviors that
emerge from the collective actions of many drivers, like the propagation of
traffic disturbances such as tailbacks in the opposite direction to traffic
flow, and the so-called jamming transition, where cars' speeds drop
suddenly as traffic density passes a critical point.  Some similar
phenomena are visible in pedestrian traffic as well, although pedestrians
are not always confined to a one-dimensional road the way cars are, and the
added freedom can give rise to additional phenomena.
\begin{bibblock}
\bibnum{NS92}\quad The classic Nagel--Schreckenberg model
  of road traffic is a beautiful example of the application of now-standard
  ideas from complex systems theory to a real-world problem.  The model is
  a ``cellular automaton'' model.  Cellular automata are discussed further
  in Section~\ref{sec:discrete}.~(I)
\bibnum{Helbing01}\quad The Nagel--Schreckenberg model and
many other models and theories of traffic flow are examined in detail in
this extensive review by Helbing.~(I)
\end{bibblock}
Flocking or schooling in birds or fish is a cooperative phenomenon in which
the animals in a flock or school collectively fly or swim in roughly the
same direction, possibly turning as a unit.  It is believed that animals
achieve this by simple self-enforced rules that involve copying the actions
of their nearby neighbors while at the same time keeping a safe distance.
\begin{bibblock}
\bibnum{Vicsek95}\quad This paper introduces what is now the best studied
  model of flocking behavior, and a good example of a drastic but useful
  simplification of a complex problem.~(A)

\bibnum{VZ10}\quad This substantial review summarizes progress on theories
  of flocking.~(I)

\bibnum{CKFL05}\quad Another good example of the use of a simplified
  model to shed light on a complex phenomenon, this paper shows how the
  coordinated movement of a large group of individuals can self-organize to
  effectively achieve collective goals even when only a small fraction of
  individuals know where they are going.~(I)

\bibnum{Ballerini08}\quad An interesting recent development in the study
  of flocking is the appearance of quantitative studies of large flocks of
  real birds using video techniques.  This paper describes a collaborative
  project that brought together field studies with theories based on ideas
  from statistical and condensed matter physics.~(I)
\end{bibblock}

\section{Complex systems theory}
The remainder of this review deals with the general theory of complex
systems.  Perhaps ``general theories'' would be a better term, since
complex systems theory is not a monolithic body of knowledge.  Borrowing an
analogy from Doyne Farmer of the Santa Fe Institute, complex systems theory
is not a novel, but a series of short stories.  Whether it will one day
become integrated to form a single coherent theory is a matter of current
debate, although my belief is that it will not.

\subsection{Lattices and networks}
\label{sec:networks}
The current theories of complex systems typically envisage a large
collection of agents interacting in some specified way.  To quantify the
details of the system one must specify first its topology---who interacts
with whom---and then its dynamics---how the individual agents behave and
how they interact.

Topology is usually specified in terms of lattices or networks, and this is
one of the best developed areas of complex systems theory.  In most cases,
regular lattices need little introduction---almost everyone knows what a
chess board looks like.  Some models built on regular lattices are
considered in Section~\ref{sec:discrete}.  Most complex systems, however,
have more complicated non-regular topologies that require a more general
network framework for their representation.

Several books on the subject of networks have appeared in recent years.
The book by Watts below is at a popular level, although it contains a small
amount of mathematics.  The book by Newman is lengthy and covers many
aspects in technical detail; the book by Cohen and Havlin is shorter and
more selective.  I also list two reviews, one brief and one encyclopedic,
of research in the field, for advanced readers.
\begin{bibblock}
\bibnum{Watts03}~(E)
\bibnum{Newman10}~(I)
\bibnum{CH10}~(I)
\bibnum{Strogatz01}~(A)
\bibnum{Boccaletti06}~(A)
\end{bibblock}
The book by Easley and Kleinberg cited earlier as Ref.~\ref{bib:kleinberg}
also includes material on networks.

\subsection{Dynamical systems}
Turning to the behavior of the agents in a complex system, many different
theories have been developed.  One of the most mature is dynamical systems
theory, in which the behaviors of agents over time are represented
individually or collectively by simple mathematical models, coupled
together to represent interactions.  Dynamical systems theory is divided
into continuous dynamics, addressed in this section, and discrete dynamics,
addressed in the following one.

Continuous dynamical systems are typically modeled using differential
equations and show a number of emergent behaviors that are characteristic
of complex systems, such as chaos and bifurcations (colorfully referred to
as ``catastrophes'' in the 1970s, although this nomenclature has fallen out
of favor).  Three elementary references are the following:
\begin{bibblock}
\bibnum{Strogatz03}\quad A popular book introducing some of the basic ideas
of dynamical systems theory by one of the pioneers of the field.  The book
focuses particularly on the phenomenon of synchronization, but also
includes useful material on other topics.~(E)
\bibnum{PJS04}\quad A lavishly illustrated introduction suitable for
undergraduates or even advanced high-school students.~(E)
\bibnum{AS92}\quad This unusual book is, sadly, out of print now, though
one can still find it in libraries.  It is essentially a picture book or
comic illustrating the principles of dynamical systems.  The field being
one that lends itself well to visual representation, this turns out to be
an excellent way to grasp many of the basic ideas.~(E)
\end{bibblock}
There are also many more advanced sources for material on dynamical
systems, including the following.
\begin{bibblock}
\bibnum{Strogatz94}\label{bib:strogatz}\quad A substantial college-level
  text on the standard methods of dynamical systems theory.~(I)
\bibnum{Lorenz63}\quad This is a classic in the field, the first paper to
  really spell out the origin of chaotic behavior in a simple system, and
  is clear and well written, although it requires a strong mathematical
  background.~(A)
\bibnum{OGY90}\quad Another seminal paper in the field, which studies the
  technically important subject of controlling chaotic systems.~(A)
\end{bibblock}

\subsection{Discrete dynamics and cellular automata}
\label{sec:discrete}
Discrete dynamical systems, those whose evolution in time progresses via a
succession of discrete ``time steps,'' were a subject of considerable
research interest in the 1970s and 1980s.  A classic example is the
logistic map, which displays a transition (actually several transitions)
from an ordered regime to a chaotic one that inspired a substantial
literature on the ``edge of chaos'' in complex systems.
\begin{bibblock}
  \bibnum{May76}\quad A classic pedagogical review of the logistic map and
  similar discrete dynamical systems from one of the fathers of complex
  systems theory.  The mathematics is elementary in principle, involving
  only algebra and no calculus, but some of the concepts are nonetheless
  quite tricky to visualize.~(I)
  \bibnum{Feigenbaum83}\label{bib:feigenbaum}\quad In 1978 Mitchell
  Feigenbaum proved one of the most important results in dynamical systems
  theory, the existence of universal behavior at the transition to chaos,
  deriving in the process a value for the quantity now known as
  Feigenbaum's constant.  His original research papers on the topic are
  technically challenging, but this later paper is relatively approachable
  and provides a good outline of the theory.~(I)
\end{bibblock}
A pedagogical discussion of Feigenbaum's theory can also be found in the
book by Strogatz, Ref.~\ref{bib:strogatz} above.

Dynamical systems that are discrete in both time and space are called
\defn{cellular automata}, or CAs for short, and these fall squarely in the
realm of complex systems, being precisely systems of many interacting
agents.  The simplest and best studied cases are on lattices, although
cellular automata with other geometries are also studied.  Well known
examples of cellular automata include J.~H.~Conway's ``Game of Life,'' the
``Rule 110'' automaton, which is capable of universal computation, and the
Nagel--Schreckenberg traffic model mentioned in
Section~\ref{sec:individual}.
\begin{bibblock}
\bibnum{Gardner70}\quad One of Martin Gardner's excellent ``Mathematical
  Games'' columns for \textit{Scientific American}, in which the most
  famous CA of them all, Conway's Game of Life, made its first appearance.
  Decades later the article is still an excellent introduction.~(E)

\bibnum{BCG03}\quad This is the second of four excellent volumes about
  games---such as board games and card games---and their mathematical
  analysis, originally published in the 1980s but recently republished.  It
  contains a thorough discussion of the Game of Life, which was invented by
  one of the book's authors.~(I)

\bibnum{Dennett98}\quad This book is not, principally, a book about CAs
  and its author is not principally a CA researcher, but the chapter
  entitled ``Real Patterns'' is an excellent introduction not only to CAs
  but also to why those who study complex systems are interested in them as
  models of processes in the wider world.~(E)

\bibnum{Wolfram02}\quad Most of this large volume is devoted to a
  discussion of the author's research, but the first part of the book,
  particularly the first hundred pages or so, provides a very readable
  introduction to CAs, laying out the basics of the field clearly while
  making only modest mathematical demands of the reader.~(I)

\bibnum{Langton86}\quad An influential early paper on the theory of
  cellular automata, which made connections with other areas of complex
  systems research, including chaos theory and ``artificial life'' (see
  Section~\ref{sec:abm}).  Among other things, the paper contains some (in
  retrospect) rather charming figures of simulation results, created by
  directly photographing the screen of a computer terminal.~(I)

\bibnum{Ilachinski01}\quad For the advanced reader this book provides
  most of what one might want to know about cellular automata.~(A)
\end{bibblock}
Chapter~11 of the book by Mitchell, Ref.~\ref{bib:mitchell}, also provides
a good overview of the study of cellular automata.  For those interested in
pursuing the topic further, an excellent and entertaining resource is the
free computer program \textit{Golly}, by Andrew Trevorrow and Tomas
Rokicki, which simulates a wide range of cellular automata and illustrates
their dynamics with instructive and elegant computer graphics.

\subsection{Scaling and criticality}
\label{sec:scaling}
Among the fundamental tools in the theory of complex systems, some of the
most important have been the physical ideas of scaling, phase transitions,
and critical phenomena.  One example of their application is mentioned
above, the study by Feigenbaum of critical behavior in discrete dynamical
systems at the ``edge of chaos,'' Ref.~\ref{bib:feigenbaum}, but there are
many others.

A startling phenomenon observed in a number of complex systems is the
appearance of ``power-law'' distributions of measured quantities.
Power-law distributions are said to ``scale'' or ``show scaling'' because
they retain their shape even when the measured quantity is ``rescaled,''
meaning it is multiplied by a constant.  The observation and origin of
power laws and scaling in complex systems has been a subject of discussion
and research for many decades.  The following two papers provide general
overviews of the area:
\begin{bibblock}
\bibnum{Mitzenmacher04}~(I)
\bibnum{Newman05b}~(I)
\end{bibblock}
Power laws have been the topic of some of the most influential publications
in complex systems theory, going back as far as the work of Pareto in the
1890s.  The mechanisms for power-law behavior have been a particular focus
of interest and the claim has been made that there may be a single
mathematical mechanism responsible for all power laws and hence a unified
theory of complex systems that can be built around that mechanism.  One
candidate for such a universal mechanism is ``self-organized criticality.''
Current thinking, however, is that there are a number of different
mechanisms for power-law behavior, and that a unified theory probably does
not exist.
\begin{bibblock}
\bibnum{Simon55}\quad One of the first, and still most important,
  mechanisms suggested for power laws, is the ``rich get richer'' or
  ``preferential attachment'' mechanism.  Simon was the first to write down
  the theory in its modern form, although many of the ideas were present in
  significantly earlier work: see for instance \bibentry{Yule25}.~(A)

\bibnum{BTW87}\quad Physicists have long been aware that physical systems
  tuned precisely to a special ``critical point'' will display power-law
  behavior, but on its own this appears to be a poor explanation for power
  laws in naturally occurring complex systems, since such systems will not
  normally be tuned to the critical point.  Bak~\etal\ in this paper
  proposed an ingenious way around this problem, pointing out that certain
  classes of system tune themselves to the critical point automatically,
  simply by the nature of their dynamics.  This process, dubbed
  ``self-organized criticality,'' is illustrated in this paper with a
  cellular automaton model, the ``self-organizing sandpile.''~(A)

\bibnum{KF86}\quad Sometimes overlooked in the literature on
  self-organized criticality, this paper actually preceded the paper by
  Bak~\etal\ by more than a year and described many of the important
  concepts that formed the basis for the approach of Bak~\etal~(A)

\bibnum{DS92}\quad Perhaps the simplest of self-organized critical
  models is the forest fire model of Drossel and Schwabl.  Although it came
  after the sandpile model of Bak~\etal\ it is easier to understand and may
  make a better starting point for studying the theory.~(A)

\bibnum{Bak96}\quad A self-contained and readable, if somewhat partisan,
  introduction to the science of self-organized criticality, written by the
  theory's greatest champion.~(E)

\bibnum{CD99}\quad An alternative general theory for the appearance of
  power laws is the ``highly optimized tolerance'' (HOT) theory of Carlson
  and Doyle.  While its inventors would not claim it as an explanation of
  all power laws, it may well be a better fit to observations than
  self-organized criticality in some cases.  This paper introduces the
  best-known model in the HOT class, the ``highly optimized forest fire''
  model, which is analogous to the self-organized forest fire model
  above.~(I)

\bibnum{WBE97}\quad Perhaps the biggest stir in this area in recent years
  has been created by the theory of biological allometry, i.e.,~power-law
  scaling in biological organisms, put forward by West~\textit{et~al.}
  This is the original paper on the theory, although West~\etal\ have
  published many others since.~(A)

\bibnum{WB04}\quad A general introduction to the theory of West~\etal\
  for physicists.~(E)
\end{bibblock}
The book by Mandelbrot, Ref.~\ref{bib:mandelbrot}, is also an important
historical reference on this topic, making a connection between power laws
and the study of fractals---curves and shapes having non-integer dimension.

\subsection{Adaptation and game theory}
\label{sec:gametheory}
A common property of many though not all complex systems is adaptation,
meaning that the collective behavior of the agents in the system results in
the optimization of some feature or quantity.  Biological evolution by
means of natural selection is the classic example: evolution takes place as
a result of the competition among the members of a breeding population for
resources and is thus exclusively a result of agent
interactions---precisely an emergent phenomenon in the complex systems
sense.

Complex systems displaying adaptation are sometimes called ``complex
adaptive systems.''  In constructing theories and models of complex
adaptive systems the fundamental concept is that of ``fitness,'' a measure
or value that conveys how well an individual, group, species, or strategy
is doing in comparison to the competition, and hence how likely it is to
thrive.  In the simplest models, one posits a fitness function that maps
descriptive parameters, such as body size or foraging strategy, to fitness
values and then looks for parameter values that maximize the fitness.

The following three books are not specifically about complex systems, but
nonetheless all provide an excellent background for the reader interested
in theories of adaptation.
\begin{bibblock}
\bibnum{MaynardSmith93}\quad This updated version of
  Maynard Smith's widely read introduction to evolutionary theory is still
  a good starting point for those who want to know the basics.~(I)

\bibnum{Dawkins97}\quad Dawkins is one of the best known science writers
  of the last century and his many books on evolutionary biology have been
  particularly influential.  His earlier book \textit{The Selfish Gene} is,
  after Darwin's \textit{Origin of Species}, perhaps the most influential
  book written about evolution.  \textit{Climbing Mount Improbable} is more
  elementary and, for the beginner, an excellent introduction to our
  current understanding of the subject.~(E)

\bibnum{Gould02}~(I)
\end{bibblock}

Biologically derived ideas concerning adaptation have also inspired
applications in computer science, wherein practitioners arrange for
programs or formulas to compete against one another to solve a problem, the
winners being rewarded with ``offspring'' in the next generation that then
compete again.  Over a series of generations one can use this process to
evolve good solutions to difficult problems.  The resulting method, under
the names \defn{genetic algorithms} or \defn{genetic programming}, has
become a widely used optimization scheme and a frequent tool of complex
systems researchers.
\begin{bibblock}
\bibnum{Holland92}\quad A nontechnical introduction to
genetic algorithms by their originator and greatest proponent, John
Holland.~(E)
\bibnum{Koza03}\quad A discussion of genetic programming,
which is the application of genetic-algorithm-type methods directly to the
evolution of computer software.~(E)
\bibnum{Mitchell96}\quad Although relatively old, Mitchell's book
on genetic algorithms is probably still the foremost general text on the
subject and a good resource for those looking for more depth.~(A)
\end{bibblock}
While fitness can depend on simple physical parameters like body size,
significant contributions to fitness at the organismal level often come
from the behaviors of agents---the way they interact with each other and
their environment.  The mapping between the parameters of behavior and the
fitness is typically a complex one and a body of theory has grown up to
shed light on it.  This body of theory goes under the name of \defn{game
  theory}.

A ``game,'' in this sense, is any scenario in which ``players'' choose from
a set of possible moves and then receive scores or ``payoffs'' based on the
particular choice of moves they and the other players made.  Game theory is
used in the context of biological evolution to model mating strategies, in
economics as a model of the behavior of traders in markets, in sociology to
model individuals' personal, financial, and career decisions, and in a host
of other areas ranging from ecology and political science to computer
science and engineering.

Although almost a quarter of a century old, Morton Davis's ``nontechnical
introduction'' to game theory remains a good starting point for those
interested in understanding the ideas of game theory without getting into a
lot of mathematics.  The book has been recently reprinted in an inexpensive
paperback edition that makes it a good buy for students and researchers
alike.  For a more mathematical introduction, the book by Myerson is a
classic, written by one of the leading researchers in the field, while the
book by Watson gives a lucid modern presentation of the material.
\begin{bibblock}
\bibnum{Davis97}~(E)
\bibnum{Myerson97}~(A)
\bibnum{Watson07}~(I)
\end{bibblock}
The book by Nowak, Ref.~\ref{bib:nowak}, also provides an introduction to
game theoretical methods specifically in the area of biological evolution,
while the book by Easley and Kleinberg, Ref.~\ref{bib:kleinberg}, includes
a discussion of games played on networks.

Some specific topics within game theory are so important and widely
discussed that a knowledge of them is a must for anyone interested in
the area.
\begin{bibblock}
\bibnum{Axelrod06}\quad The ``prisoner's dilemma'' is probably the best
  known (and also one of the simplest) of game theoretical examples.  A
  famous event in the history of game theory is the contest organized by
  Robert Axelrod in which contestants devised and submitted strategies for
  playing the (iterated) prisoner's dilemma game against one another.
  Among a field of inventive entries, the contest was won by mathematical
  biologist Anatol Rapoport using an incredibly simple strategy called
  ``tit-for-tat,'' in which on each round of the game the player always
  plays the same move their opponent played on the previous round.  Axelrod
  uses this result as a starting point to explain why people and animals
  will sometimes cooperate with one another even when it is, at first
  sight, not in their own best interests.~(E)

\bibnum{CZ97}\quad The minority game, proposed by physicists Challet and
  Zhang, is a remarkably simple game that nonetheless shows complex and
  intriguing behavior.  In this game a population of $n$ players, where $n$
  is odd, repeatedly choose one of two alternative moves, move 1 or move~2.
  On any one round of the game you win if your choice is in the minority,
  i.e.,~if fewer players choose the same move as you than choose the
  alternative.  It's clear that there is no universal best strategy for
  playing this game since if there were everyone would play it, and then
  they'd all be in the majority and would lose.  The minority game is a
  simplified version of an earlier game proposed by Brian Arthur, usually
  called the \defn{El~Farol problem}, in honor of a famous bar of that name
  in Santa Fe, New Mexico.~(A)

\bibnum{Gintis09}\quad An intriguing line of work in the last couple of
  decades has been the development of experimental game theory (also called
  behavioral game theory or experimental economics), in which instead of
  analyzing games theoretically, experimenters get real people to play them
  and record the results.  The remarkable finding is that, although for
  many of these games it is simple to determine the best move---even
  without any mathematics---people often don't play the best move.  Even if
  the experimenters offer real money in return for winning plays, people
  routinely fail to comprehend the best strategy.  Results of this kind
  form the basis for the economic theory of ``bounded rationality,'' which
  holds that it is not always correct to assume that people act in their
  own best interests with full knowledge of the consequences of their
  actions.  (This may seem like an obvious statement, but it is a
  surprisingly controversial point in economics.)~(A)
\end{bibblock}

\subsection{Information theory}
Information theory is not usually regarded as a part of complex systems
theory itself, but it is one of the tools most frequently used to analyze
and understand complex systems.  As its name suggests, information theory
describes and quantifies information and was originally developed within
engineering as a way to understand the capabilities and limitations of
electronic communications.  It has found much wider application in recent
years, however, including applications to the analysis of patterns of many
kinds.  A pattern is precisely recognizable as a pattern because its
information content is \emph{low}.  For instance, there is little
information in a periodically repeating sequence of symbols, numbers,
colors, etc.  If we can accurately predict the next symbol in a sequence
then that symbol contains little information since we knew what it was
going to be before we saw it.  This idea and its extensions have been
applied to the detection of patterns in DNA, in networks, in dynamical
systems, on the Internet, and in many kinds of experimental data.
\begin{bibblock}
\bibnum{Pierce80}\quad Although relatively old, this book is still the
  best introduction to information theory for the beginner.  The subject
  requires some mathematics for its comprehension, but the level of
  mathematical development in Pierce's book is quite modest.~(I)

\bibnum{CT91}\quad A thorough introduction to modern information theory,
  this book demands some mathematical sophistication of the reader.~(A)

\bibnum{Shannon48a}\quad The original paper by the father of information
  theory, Claude Shannon, in which he lays out the theory, in remarkably
  complete form, for the first time.  As well as being the first paper on
  the topic, this is also a well-written and palatable introduction for
  those willing to work through the mathematics.~(A)
\end{bibblock}
An active area of current research in complex systems is the application of
information theory to measure the complexity of a system.  This work aims
to answer quantitatively the question, ``What is a complex system?''\ by
creating a measure that will, for instance, take a large value when a
system is complex and a small one when it is not.  One of the best-known
examples of such a measure is the Kolmogorov complexity, which is defined
as the length of the shortest computer program (in some agreed-upon
language) that will generate the system of interest or a complete
description of it.  If a system is simple to describe then a short program
will suffice and the Kolmogorov complexity is low.  If a larger program is
required then the complexity is higher.  Unfortunately the Kolmogorov
complexity is usually extremely hard---and in some cases provably
impossible---to calculate, and hence researchers have spent considerable
effort to find measures that are more tractable.
\begin{bibblock}
\bibnum{Bennett90}\quad A nontechnical description of the problem and
why it is interesting by one of the leading researchers in the field.~(E)
\bibnum{BP97}\quad Chapters 8 and 9 of this book provide a useful
introduction to measures of complexity, and provide a connection to the
topic of the next section of this review, computational complexity
theory.~(A)
\end{bibblock}

\subsection{Computational complexity}
Somewhat peripheral to the main thrusts of current complex systems
research, but nonetheless of significant practical value, is the study of
\defn{computational complexity}.  Computational complexity theory deals
with the difficulty of performing certain tasks, such as calculating a
particular number or solving a quantitative problem.  Although typically
discussed in the language of algorithms and computer science, computational
complexity in fact has much wider applications, in evolutionary biology,
molecular biology, statistical physics, game theory, engineering, and other
areas.  For instance, one might ask how difficult is it, in terms of time
taken or number of arithmetic computations performed, to the find the
ground state of a physical system, meaning the state with the lowest
energy, among all possible states.  For some systems this is an easy task
but for others it is difficult because there are many possible states and
no general principle for predicting which will have the lowest energy.
Indeed it is possible to prove, subject to basic assumptions, that in some
cases there exists no general technique that will find the ground state
quickly, and the only reliable approach is to search exhaustively through
every state in turn, of which there may be a huge number.  But if this is
true for computations performed by hand or on a computer, it is no less
true of nature itself.  When nature finds the lowest energy state of a
system it is, in effect, performing a computation, and if you can prove
that no method exists for doing that computation quickly then this tells
you that the physical system will not reach its ground state quickly, or in
some cases at all, if the number of states that need to be searched through
is so vast that the search would take years or centuries.  Thus results
about the theory of computation turn out to give us very real insight into
how physical (or biological or social) systems must behave.

The best known issue in computational complexity theory, one that has made
it to the pages of the newspapers on occasion, is the question of whether
two fundamental classes of problems known as P and NP are in fact
identical.  The class~P is the class of problems that can be solved
rapidly, according to a certain definition of ``rapidly.''  An example is
the problem of multiplying two matrices, for which there is a simple
well-known procedure that will give you the answer in short order.  The
class~NP, on the other hand, is the class of problems such that if I hand
you the solution you can \emph{check that it's correct} rapidly, which is
not the same thing at all.  Obviously NP includes all problems in~P---if
you tell me a purported solution for the product of two matrices I can
check it rapidly just by calculating the product myself from scratch and
making sure I agree with your answer.  But NP can also include problems
whose answer is easy to check but difficult to compute.  A classic example
is the ``traveling salesman problem,'' which asks whether there exists a
route that will take a salesman to each of $n$ cities while traveling no
more than a set number of miles.  (It is assumed, for simplicity, that the
salesman can fly in a straight line from each city on his route to the
next---he is not obliged to follow the path of the established roads.)  If
you hand me a purported solution to such a problem I can check it quickly.
Does the route visit every city?  Is it below the given number of miles?
If the answer is yes to both questions then the solution is good.  But if
you give me only the list of cities and I have to find a solution for
myself then the problem is much harder because there are many possible
routes to consider and no general rule for predicting which ones will meet
the specified length limit.  Indeed it widely is believed (though not known
for certain) that no method exists that will find the solution to the
traveling salesman problem rapidly in all cases; there is nothing you can
do that is much better than simply going through all routes one by one
until you find one that works.

Unless this belief is wrong and there exists a (currently unknown) way to
solve such problems easily so that problems in the NP class also belong
to~P, then NP is a bigger class than P and hence the two classes are not
identical.  Most researchers in computational complexity theory believe
this to be the case, but no one has yet been able to prove it, nor indeed
has any clue about how one should even begin.
\begin{bibblock}
\bibnum{Aaronson05}\quad In this article Aaronson discusses
  the application of computational complexity theory, and particularly the
  central idea of ``NP-completeness,'' to a wide range of scientific
  problems including protein folding, quantum computing, and relativity,
  introducing in the process many of the main ideas of computational
  complexity.~(I)

\bibnum{MM11}\quad A readable and informative introduction to the theory
  of computational complexity and its applications from two leading complex
  systems researchers.  This book emphasizes the important idea that it is
  not only computers that perform computation: all sorts of systems in the
  natural and man-made world are effectively performing computations as
  part of their normal functioning, and so can be viewed through the lens
  of computational theories.~(I)

\bibnum{Sipser06}\quad A general and widely used text on computational
  complexity within computer science.~(A)
\end{bibblock}

\subsection{Agent-based modeling}
\label{sec:abm}
Many types of computer modeling are used to study complex systems.  Most of
the standard methods of numerical analysis---integration methods, linear
algebra and spectral methods, Monte Carlo methods, and so forth---have been
applied in one branch of the field or another.  However, there is one
method that is particular to the study of complex systems and has largely
been developed by complex systems scientists, and that is \defn{agent-based
  modeling}.  The goal of agent-based computer models, sometimes also
called ``individual-based,'' is to separately and individually simulate the
agents in a complex system and their interactions, allowing the emergent
behaviors of the system to appear naturally, rather than putting them in by
hand.  The first two papers listed here both give pedagogical introductions
to agent-based methods, but from quite different viewpoints.  The third
reference is an entire journal volume devoted to discussions of agent-based
modeling, including a number of accessible overview articles.
\begin{bibblock}
\bibnum{Page08}~(E)
\bibnum{MW02}~(E)
\bibnum{PNAS02}~(E)
\end{bibblock}
The book by Miller and Page, Ref.~\ref{bib:MP}, also contains a useful
introduction to agent-based methods.  There also exist a number of books
that tackle the subject in the context of specific fields of scientific
study, such as:
\begin{bibblock}
\bibnum{GR05}\quad An introduction to agent-based modeling in ecology.~(I)
\bibnum{Gilbert07}\quad A very short introduction to social science
applications of agent-based models.~(I)
\end{bibblock}
A few classic examples of agent-based models are also worthy of mention:
\begin{bibblock}
\bibnum{Schelling71}\quad One of the first true agent-based models is the
  model of racial segregation proposed by Thomas Schelling in 1971.
  Schelling did not have access to a computer at the time he proposed his
  model (or perhaps was not interested in using one), and so simulated it
  by hand, using coins on a grid of squares.  However, many computer
  simulations of the model have subsequently been performed.  Schelling was
  awarded the Nobel Prize in Economics in 2005, in part for this work, and
  to date this is the only Nobel Prize awarded for work on traditional
  complex systems (although one could argue that, for instance, condensed
  matter systems are complex systems, and several prizes in physics have
  been awarded for condensed matter research).~(E)

\bibnum{EA96}\quad The ``Sugarscape'' models of Epstein and Axtell
  provide a beautiful example of the emergence of complex behaviors from
  the interactions of simple agents.  This set of models would also be a
  good starting point for experimenting with agent-based simulations: the
  rules are simple and easy to implement, and the results lend themselves
  nicely to computer graphics and visualization, making the models
  relatively straightforward to interpret.  Versions of some of the models
  are available already programmed in standard agent-based simulation
  software packages (see below).~(I)

\bibnum{Palmer94}\quad A good example of an agent-based model is the
  ``artificial stock market'' created by Palmer~\etal\ at the Santa Fe
  Institute in the early 1990s.  In this study, the researchers simulated
  individually the behavior of many traders in a stock market, giving them
  a deliberately heterogeneous selection of trading strategies and limited
  knowledge of market conditions.  They observed regimes of the model in
  which it displayed the equilibrium behavior of neoclassical economics,
  but others in which it displayed chaotic behavior more akin to that of
  real stock markets.~(I)

\bibnum{Ray91}\quad An inventive and influential example of an
  agent-based simulation is the Tierra evolution model created by Ray.  In
  this simulation, computer programs reproduce by explicitly copying
  themselves into new memory locations, competing and mutating to make best
  use of computer resources, meaning CPU time and memory.  Although similar
  in some respects to the genetic programming studies discussed in
  Section~\ref{sec:gametheory}, Tierra is different in that no fitness
  function is imposed externally upon its programs.  Instead, fitness
  emerges naturally in the same way it does in biological evolution: those
  programs that manage to reproduce themselves survive and spread, while
  those that do not die out.  Tierra was the first such simulation to be
  constructed, but others, such as the Avida system, have appeared in
  recent years.  Systems such as these are referred to generically as
  ``artificial life'' simulations.  Artificial life was a major thrust in
  complex systems research in the 1990s.~(I)
\end{bibblock}
Finally, there are a variety of software packages available for performing
agent-based simulations.  Some of them are highly advanced programming
libraries suitable for cutting-edge research, while others are designed as
easy-to-use educational tools requiring little prior knowledge.  Among the
former, \textit{Repast} and \textit{Mason} are currently the most widely
used and mature systems, while among the latter \textit{NetLogo} is a good
starting point.

\section{Conclusion}
Complex systems is a broad field, encompassing a wide range of methods and
having an equally wide range of applications.  The resources reviewed here
cover only a fraction of this rich and active field of study.  For the
interested reader there is an abundance of further resources to be explored
when those in this article are exhausted, and for the scientist intrigued
by the questions raised there are ample opportunities to contribute.
Science has only just begun to tackle the questions raised by the study of
complex systems and the areas of our ignorance far outnumber the areas of
our expertise.  For the scientist looking for profound and important
questions to work on, complex systems offers a wealth of possibilities.

\begin{acknowledgments}
  I thank Doyne Farmer, Rick Riolo, Roger Stuewer, and several anonymous
  referees for useful comments and suggestions.  This work was funded in
  part by the James S. McDonnell Foundation.
\end{acknowledgments}

\end{document}